\documentclass{article}

\usepackage{threeparttable}
\usepackage{graphicx}
\usepackage{float}
\graphicspath{ {figures/} }

\title{To What Extent Are Honeypots and Honeynets Autonomic Computing Systems?}

\author{Jason M. Pittman, Shaho Alaee}

\begin{document}
	\maketitle

	\section{Abstract}
	Cyber threats, such as advanced persistent threats (APTs), ransomware, and zero-day exploits, are rapidly evolving and demand improved security measures. Honeypots and honeynets, as deceptive systems, offer valuable insights into attacker behavior, helping researchers and practitioners develop innovative defense strategies and enhance detection mechanisms. However, their deployment involves significant maintenance and overhead expenses. At the same time, the complexity of modern computing has prompted the rise of autonomic computing, aiming for systems that can operate without human intervention. Recent honeypot and honeynet research claims to incorporate autonomic computing principles, often using terms like adaptive, dynamic, intelligent, and learning. This study investigates such claims by measuring the extent to which autonomic principles principles are expressed in honeypot and honeynet literature. The findings reveal that autonomic computing keywords are present in the literature sample, suggesting an evolution from self-adaptation to autonomic computing implementations. Yet, despite these findings, the analysis also shows low frequencies of self-configuration, self-healing, and self-protection keywords. Interestingly, self-optimization appeared prominently in the literature. While this study presents a foundation for the convergence of autonomic computing and deceptive systems, future research could explore technical implementations in sample articles and test them for autonomic behavior. Additionally, investigations into the design and implementation of individual autonomic computing principles in honeypots and determining the necessary ratio of these principles for a system to exhibit autonomic behavior could provide valuable insights for both researchers and practitioners.

	\section{Introduction}
	
	One may wonder why honeypots and honeynets exist as deception technologies. Anecdotally, honeypots and honeynets are often discussed but rarely encountered in the wild. Without any doubt, the increasing sophistication of cyber threats, including advanced persistent threats (APTs), ransomware, and zero-day exploits, demand improved security measures to protect the enterprise. Honeypots and honeynets can contribute to the development of innovative defense strategies. Such \textit{deceptive systems} do so by allowing researchers and practitioners to capture malicious behavior. Captured behavior allows for development of new detection  mechanisms and deeper understanding of evolving threats.

	Indeed, gaining insights into attacker tactics, techniques, and procedures (TTPs) is crucial for developing effective countermeasures. Honeypots and honeynets serve as controlled environments that allow researchers to study attacker behavior, collect valuable data on their strategies, and identify trends in cyberattacks. Furthermore, honeypots and honeynets can serve as testbeds for developing, evaluating, and refining new security technologies, such as machine learning algorithms, intrusion detection systems, and automated response mechanisms.
	
	Unfortunately, honeypots and honeynets have high maintenance and overhead costs \cite{Bringer_2012, Campbell_2015, Ikuomenisan_2022}. After all, honeypots and honeynets are only as good as the malicious traffic the system attracts. Consequently, keeping these systems configured correctly, rebuilding after compromises, and adapting services to changes in operating environments are necessary. Much of that work is manual or, at best, partially automated \cite{Gupta_2018}.
	
	Honeypots and honeynets are not the only modern computing systems to suffer from such problems, though. The complexity of modern computing has risen to a level whereby there is a strong incentive to develop the systems that can operate without human involvement. In fact, the field of autonomic computing \cite{horn2001autonomic} seeks to do so in a general case. Meanwhile, in a specific case, a variety of studies have been published in which some degree of autonomic computing is expressed for honeypots and honeynets. However, it is not clear to what extent honeypots and honeynets implement true autonomic computing principles. 
	
	This work provides a systematic review of state-of-the-art honeypot and honeynet literature through the lens of autonomic computing. The deceptive system literature uses a handful of functional labels to describe cutting edge honeypots and honeynets (e.g., intelligent, adaptive, dynamic, and so forth).  Thus, one goal of this work is to establish operational definitions for these labels. Relatedly, another goal is to establish a semantic mapping between deceptive system labels and the principles of autonomic computing.   
	
	The rest of this work is organized in four sections. First, we describe a robust background literature consisting of relevant autonomic computing research. This is followed by a systematic treatment of state-of-the-art honeypot and honeynet research. Together, these sections establish the conceptual framework necessary to operationalize the semantics associated with deceptive systems and autonomic computing. Then, in the next section, we describe the method used to investigate label definitions and to construct the semantic mapping. The results of the investigation and construction are then presented using both qualitative and quantitative measures. Finally, we offer recommendations and ideas for follow-on research as part of the conclusions.
	
	\section{Related Work}
	The following research background is not intended to be exhaustive. Rather, our intention is to establish a semantic foundation for the convergence of autonomic computing, autonomous intelligent systems, and honeypots or honeynets. Thus, the following sections highlight seminal literature containing definitive terms with clear operational meaning.
	
	\subsection{Autonomic Computing Principles}
	The complexities of computing systems has led to a rise in maintenance and operations workload that surpasses the capacity of human capabilities \cite{psaier2011survey}. Automation has long been a technique employed to delay the crushing weight of these workloads. Yet, even automation cannot keep up with the increases in distribution of compute decision-making such as Edge, IoT, and multi-cloud \cite{choudhury2021autonomic}. 
	
	The field of autonomic computing seeks to develop systems capable of self-management \cite{horn2001autonomic, Parashar2004} or self-governing automation. Such systems operate without explicit human intervention \cite{horn2001autonomic, Dehraj2021} and do not require training. Further, autonomic computing systems do not require external administration or maintenance. These are broad concepts though and the semantic needs to be clarified.
	
	Autonomic computing is not to be confused with autonomous systems. The latter are generally designed to perform a specific task or set of tasks, while autonomic computing refers to the overall ability of a system to manage itself and its resources.  Moreover, it is worth noting autonomous systems can exhibit autonomous intelligence, but not all autonomous systems are intelligent. The differences here are emphasized by the core principles of autonomic computing.
	
	Self-management is realized when it exhibits self-configuring, self-healing, self-optimization, and self-protection behaviors \cite{horn2001autonomic, kephart2003}. More specifically, the system can automatically configure itself based on its environment or workload. The system can detect and correct errors or failures. Additionally, the system seeks equilibrium between its performance or resource utilization and any given task. Lastly, the system is capable of defending itself from protect threats. Importantly, these are not reducible concepts. Further, while each may be developed and implemented in isolation, there is a moderate coupling between each at the system level. Certainly, all principles must be implemented for a system to be considered \textit{autonomic}.
	
	More recently, the literature has begun to refer in the collective to the four principles as \textit{self-adaptation} \cite{Henrichs2022, Quin2022, Wong2022}. This semantic drift becomes critical for conceptualizing our related work as we move to honeypots and honeynets. At the same time, a salient point to consider within the context of honeypots and honeynets is whether the technology implements autonomic computing principles or the technology enables autonomic computing principles in adjacent systems. The directional difference between these two options speaks to the semantic relationship between general autonomic computing research versus specific implementations of the paradigm. 

	\subsection{Honeypots and Honeynets}
	Honeypots and honeynets are designed to be targeted, investigated, or breached with the aim of gathering information about malicious activities \cite{curran2005monitoring}. Researchers and experts can analyze the patterns and actions \cite{curran2005monitoring, Campbell_2015} by engineering such systems to be intentionally vulnerable. As a deceptive technologies, both honeypots and honeynets rest upon a rich literature and have been extensively mapped in regards to critical semantics \cite{Bringer_2012, Zakaria_2013, Mohan_2022}. 
	
	Briefly, there are a variety of types and architectures associated with honeypots and honeynets. Categorization has traditionally been according to type, deployment form, or more commonly by interaction \cite{Zakaria_2013, Campbell_2015}. Low and high interaction are the most common types by interaction and refer to whether the honey system exposes a minimal functionality (i.e., low) or a full application service or operating system (i.e., high). While early honeypots and honeynets consisted of physical computing systems and networks \cite{Mohammadzadeh_2013}, modern deception implementations are virtualized or containerized and cloud-capable \cite{Gupta_2018}. Additionally, honeypot inspired deceptive systems have evolved in lockstep with emerging technologies such as software-defined networking, tokenized services, or \textit{as-a-service} models \cite{Han_2016, Gupta_2018, Ikuomenisan_2022}.
	
	Echoing the impetus for autonomic computing research, a plethora of honeypot and honeynet researchers cite high costs of administration and maintenance as problems \cite{Hecker2012, Zakaria_2013, Fan2019}. Furthermore, the literature overtly recognizes discoverability and sojourn time \cite{Rowe2019, Pittman2020} as limitations for honeypots and honeynets. Against this background, one notable trend, or lack thereof, noted previously by Zakaria and Kiah \cite{Zakaria_2013} was the lack of innovation stemming from introducing machine learning techniques and other automation methodologies into honeypots and honeynets. While exploring research trends in honeypot and honeynet development, Ikuomenisan and Morgan \cite{Ikuomenisan_2022} found the same to be true. 
	
	The related work lens thus turns to state-of-the-art research with a focus on automation and machine learning. The following section describes the honeypot and honeynet semantic along emergent language observed in the literature. Doing so extends the types and architectures foundation into a pattern more readily analyzed in comparison to autonomic computing and autonomous intelligent systems.

	\subsection{State-of-the-Art in Honeypots and Honeynets}
	The state-of-the-art in honeypot and honeynets systems is represented in the literature through four keywords. These keywords are emergent from the language used in the research. While not strictly synonyms, the overlap in associated models and implementations suggests the research is headed along the same cone of possibility despite differences in descriptors. Adding to the potential confusion, existing literature can mix or combine multiple keywords in the same study. The next four sections should clarify the semantics important to overall understanding while preserving the original intent of the articles.
	
	Illustrative of the potential confusion, let us consider the frequency of the five keywords (Table \ref{table:frequency}). These are not exclusive frequencies, however. For example, whereas \textit{dynamic} and \textit{intelligent} represent 13 studies each, five of the studies are in the intersection of the two sets. Thus, the authors consider the honeypots and honeypots in their work to be both dynamic and intelligent. This is similarly true for other combinations as well- \textit{dynamic} and \textit{learning} exhibit an intersection of two; \textit{intelligent} and \textit{learning} and \textit{adaptive} also have an intersection of two. 
	
	\begin{table}[H]
		\centering
		\caption{Frequency of State-of-the-Art Papers}
		\begin{tabular}{ c c c c }
			\hline
			Dynamic & Intelligent & Learning & Adaptive \\ \hline \hline
			13	& 13 & 16 & 18 \\ \hline
			\label{table:frequency}
		\end{tabular}
	\end{table}
	
	The point here is not to meticulously detail a snapshot of the literature. The field is expanding and any value in such details will diminish quickly. Instead, we note the frequencies and crossovers to demonstrate the positive rising trend in the state-of-the-art literature. The meaningful details which will last are in the semantics themselves.

	\subsubsection{Dynamic}
	The initial breakaway from traditional or \textit{static} honeypots and honeynets comes in the form of dynamic honey systems. According to the research \cite{Kuwatly_2004, Hecker_2006, Park_2019}, the term \textit{dynamic} describes the ability for the honeypot or honeynet system to adapt based on changes in the system architectures surrounding it. Importantly, the honeypot or honeynet adopts the responsibility for probing the architecture detect changes. Based on predetermined rules, signatures, and configurations, the honeypot or honeynet can alter its configuration to more closely mimic newly detected architectural changes \cite{Mohammadzadeh_2013}.
	
	\subsubsection{Intelligent}
	Whereas dynamic honeypots and honeynets look at adjacent architecture, \textit{intelligent} systems examine network traffic destined for their interfaces \cite{Hecker2012, Mitchell2018}. Intelligence is closely related to the form of intelligence exhibited in artificial intelligence agents. That is, the honeypot or honeypot includes a decision-making function \cite{Han_2016, Meng_2017} which acts on information gathered through sensors. Decision outcomes are potentially more complex than what a dynamic honey system can achieve but is still constrained to the range of configurations possible given the base application or operating system. Additionally, intelligent honeypots and honeynets can compute statistical predictions and modify honey behaviors to interact with high liklihood events \cite{Naik_2020}.
	
	\subsubsection{Learning}
	The term \textit{learning} in the content of honeypots and honeynets is borrowed from the use of machine learning (ML). Naturally, a learning honeypot or honeynet extends the concepts present in the intelligent type. Similar to \textit{intelligence}, a learning honeypot or honeynet can alter its behavior based on input. Dissimilar to intelligent honeypots and honeynets however, learning-based systems examine behavior within the user environment instead of network connections \cite{Chakraborty2016, Jiang_2020}. In other words, the system learns what might maximize sojourn time and alters itself to influence the metric towards a theoretical maximal value \cite{MELLIA2021, Mohan_2022}. Strikingly, the majority of effort has gone into integrating reinforcement learning as the underlying ML technique. 
	
	\subsubsection{Adaptive}
	Where such alterations border on adaptation, the learning honey system begins to become \textit{adaptive}. In fact, adaptive honeypots and honeynets are the forward edge of state-of-the-art. Such adaptive systems leverage key aspects of dynamic, intelligent, and learning to instantiate responses to environment and behavioral stimuli \cite{Fan_2015, Dowling_2020, Landsborough2021}. These honeypots and honeynets have an expanded repertoire of behavioral and architectural changes as a result. Semantically, the research \cite{Zhang_2017, Wagener_2009, Touch_2022} often mixes and combines the four terms as a reflection of the adaptive nature in these honey systems. Yet, even adaptive honeypots and honeynets rely on human intervention at various points in their lifecycle.
	
	\subsection{Open Challenges}
	A consistent need we observed across the state-of-the-art research is for \textit{autonomy} \cite{Kuwatly_2004, Fraunholz_2017}. That is, the ability for honeypots and honeynets to operate continually without human intervention. Logically, this stands to reason given a honeypot or honeynet is a computing system and therefore all of the complexity issues driving the development of autonomic computing apply. What's more, without autonomy, the literature suggests honeypots and honeynets will not be able to escape from under the administration and maintenance burden \cite{Bringer_2012, Huang_2019, Park_2019}. Such goes hand in hand with the positive upwards trend towards ML integration we observed. Yet, ML is not autonomy per se. 
	
	\section{Method}
	
	The direction of the deceptive systems field, coupled with the semantics and terminology present in the research, implies a convergence towards honeypot and honeynet autonomic computing.  Given the four categories of honeypot and honeynet state-of-art research, we wondered if any such potential convergence is in name (i.e., lexical) only or if there is operational semantics emergent from the technical design of these systems. If we assume the implication to be true, we can pose the following research question: to what extent are autonomic computing principles expressed in forward edge honeypot and honeynet research?  
	
	A systematic review was best suited to facilitate exploring such a research question. The method involves discovering, assessing, and interpreting existing research on a specific subject \cite{Petticrew2008}. Furthermore, the systematic review process is a rigorous and structured method for synthesizing and evaluating research evidence from multiple studies \cite{Cooper1998, Kitchenham2004}. Indeed, the method aims to minimize bias, ensure transparency, and provide a comprehensive understanding of a specific research question or topic. Thus, a systematic review is appropriate when there is a well-defined research question or problem, and a substantial body of literature exists on the topic \cite{Cooper1998}. The process helps researchers identify patterns, gaps, and inconsistencies in the existing literature, making it particularly useful for informing policy, practice, or future research directions.
	
	With that in mind, we developed a research plan. The first step in the plan consisted of obtaining relevant honeypot and honeynet literature. From there, we planned to extract keywords and keyword synonyms from seminal autonomic computing literature. Extracting keywords would reveal what terms exist in the corpus and also enables creating a frequency breakdown. Then, we could measure similarity between keywords and synonyms. We could do this both with the keywords and synonyms alone as well as for each honeypot or honeynet sample paper. Colllectively, our aim is to reveal some answer to the research question. Specifically, we focused on potential lexical and semantic representation in the content, not on developing an exhaustive catalog of references. We did so in multiple phases, moving from broad to narrow in scope.
	
	We ran a search informed by the state-of-the-art using the search string \textit{self-adaptive AND (honeypot OR honeynet)}. We categorized these as \textit{self-adaptive}. Next, we sought more forward edge literature. Here, we ran a strict search using \textit{autonomic AND (honeypot OR honeynet)} as the search string. These we categorized as \textit{autonomic}. Lastly, we searched for each of four autonomic computing principles individually. We iteratively search using the string pattern of \textit{self-* principle AND (honeypot OR honeynet)}. We also included variations therein (e.g., \textit{configure} in lieu of \textit{configuration}, and so forth). 
	
	Overall, the searches yielded 166 articles. We down selected articles to a final sample of 23 (Table \ref{table:categories}) in total or 16 unique articles. Notably, four articles \cite{Pauna2018,Zarca2020,Landsborough2021,Sun2022} appear in multiple categories. Overall, the inclusion criteria consisted of a publication date in the range of 2018 to 2022, available PDF or text of the paper downloadable, and a demonstrated implementation of associated keywords. In turn, we excluded literature outside of the date range, those without an available PDF or text, or lacking a demonstrated implementation of a keyword in honeypot or honeynet.
	
	\begin{table}[H]
		\centering
		\caption{Sample Breakdown by Search Category}
		\begin{tabular}{ r c c}
			\hline
			Category & Sample & Citations \\ \hline \hline
			self-adaptive & 9 & \cite{Dowling_2020,DeFaveri2016,Huang_2019,Kashtalian2021,Landsborough2021,Niakanlahiji2020,Pauna2018,Sun2022,Suratkar2022} \\
			autonomic & 6 & \cite{feeneyhoneynet,Li2019,li2009experiment,Rowe2019,teles2011autonomic,Zarca2020} \\
			self-configuration	& 2 & \cite{Sun2022, Pauna2018} \\ 
			self-healing & 1 & \cite{Zarca2020} \\
			self-optimization & 1 & \cite{Landsborough2021} \\
			self-protection & 4 & \cite{Li2019,Zarca2020,You2021,Landsborough2021} \\
			\hline
			\label{table:categories}
		\end{tabular}
	\end{table}
	
	Additionally, we extracted a keyword list (Table \ref{table:keywords}) from two seminal autonomic computing papers \cite{horn2001autonomic,kephart2003}. The source literature did not contain honey system research, nor did the articles cite such research. Further, only a single honey system article in the sample \cite{Landsborough2021} appeared to directly cite the autonomic computing research (i.e., \cite{kephart2003}). We desired the lack of direct connection to preserve the integrity of the frequency lexical and semantic measurements as the keywords were critical as input during data analysis . 
	
	\begin{table}[H]
		\centering
		\caption{Autonomic Computing Keyword List}
		\begin{tabular}{r c}
			\hline
			Group & Keywords \\ \hline \hline
			adaptive & adapt, adaptation \\
			autonomic & autonomous, autonomy \\
			self-configuration & self-configuring, configuration, configure \\
			self-healing & self-recover, heal, recover, repair, self-repair \\
			self-optimization & self-improve, optimize, improve, performance \\
			self-protection & self-protecting, protect \\
			\hline
			\label{table:keywords}
		\end{tabular}
	\end{table}
				
	We analyzed the literature sample in two dimensions using a set of Python programs. First, we calculated descriptive statistics in the form of keyword frequencies. The aim was to uncover where the keywords appeared across the sample as well as the overall concentration levels for each keyword. We then calculated lexical similarity within each group of keywords in Table \ref{table:keywords}. Specifically, we computed Levenshtein edit distances. 
	
	Then, we measured the semantic similarity of the same keyword categories across the sample papers and produced a Mean of Means for each. We measured this using cosine similarity and a sentence embedding model. Data were collected from each paper by randomly selecting 25 sentences containing a autonomic computing keyword, calculating a Mean for the set cosine similarities for paper, and then a Mean representing the overall keyword category.  
	
	The three instruments all used a Python 3.11.2 environment. Specific packages included \textit{gensim}, \textit{nltk}, \textit{Levenshtein}, and  \textit{scikit-learn}. We used GloVe 6B \cite{Pennington2014} for the sentence embedding model during cosine similarity calculations. Common stop words were excluded through Nltk.
	
	\section{Results}
	Following from the data analysis, we organized the results according to the two dimensions of data analysis. That is, descriptive and similarity. We further organized descriptive results according to the two keyword groupings: self-adaptive and autonomic. Additionally, we present similarity as lexical and semantic. Lastly, we included one of the seminal autonomic computing papers \cite{kephart2003} as a control in both dimensions.
	
	\subsection{Descriptive Results}
	The first section of results describes what keywords appear in the honey systems literature sample. We view autonomic principles operationalized in the language of the research as features of the knowledge  as well as how we might interact with those features across fields.
		
	\subsubsection{Self-adaptive}
\begin{figure}[H]
			\includegraphics[width=\textwidth]{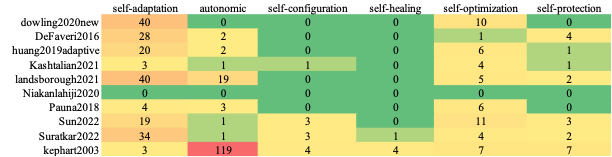}
			\caption{Frequencies of autonomic keywords in self-adaptive papers}
			\label{fig:adaptive}
	\end{figure}

	Thus, the conceptual framework demonstrated in the related work left off with \textit{adaptive}. Thus, we explored frequencies of self-adaptive category keywords first (Figure \ref{fig:adaptive}). There were $285$ keyword occurrences in total. We found a high frequency of occurrence for the \textit{self-adaptive} category ($N = 188$) excluding the control paper. The next highest frequency was the \textit{self-optimization} category ($N = 47$). The lowest frequency was found with the \textit{self-healing} category. The fewest occurrences were observed with the \textit{self-healing} keyword ($N = 1$). Meanwhile, the control paper had the highest frequency with the \textit{autonomic} category ($N = 119$) compared to $N = 29$ for the honey system articles. The control article also had frequencies in all keyword categories.
			
	\subsubsection{Autonomic}
	
\begin{figure}[H]
		\includegraphics[width=\textwidth]{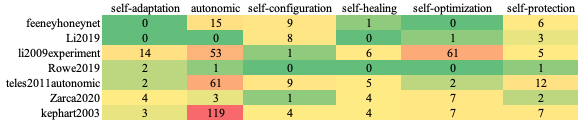}
		\caption{Frequencies of autonomic keywords in autonomic papers}
		\label{fig:autonomic}
	\end{figure}

	Pushing forward, we calculated the frequency of autonomic computing keyword appearance in the \textit{autonomic} paper category (Figure \ref{fig:autonomic}). The analysis discovered a comparable number of total keywords ($N = 299$). Moreover, we see identical results with the control study. However, the frequency of autonomic keywords in the honey system literature is different. For instance, the highest frequency of occurrence was in the \textit{autonomic} category ($N = 133$). Interestingly, the \textit{self-optimization} category had the second highest frequency again  ($N = 71$). The remaining three autonomic computing principles exhibited roughly equal frequencies. This is in contrast to the \textit{self-adaptive} category where two features held the vast majority of occurrences. As well, \textit{self-healing} again had the fewest occurrences albeit considerably more ($N = 16$).
		
	\subsection{Lexical and Semantic Similarity}
	After frequency analyses, we measured lexical and semantic similarity. The lexical similarity within each keyword group (Table \ref{table:lexical}) demonstrated to what degree each keyword within a group was similar to the category group keyword. In short, the results can be interpreted as how close the keywords are to the category roots \textit{in their use}. 
	
	To that end, we developed a Python program to measure the Levenshtein distance between keyword pairs. The edit distance was expressed as a percentage of similarity. For example, if we take the \textit{autonomic} category, we constructed two pairs as autonomic-autonomous and autonomic-autonomy. We observed the synonym \textit{autonomous} had a $70\%$ similarity to the category root (i.e., autonomic) and \textit{autonomy} had a $77\%$. Conversely, \textit{recover} has $16\%$ lexical similarity to the category root of \textit{self-healing}. 
		
	\begin{table}[H]
		\centering
		\caption{Keyword Lexical Similarity}
		\begin{threeparttable}
			\begin{tabular}{l l l l l l} 
				\hline
				Group & Similarity \\ \hline \hline  
				Adaptive & adapt & adaptation & & & \\ 
				& 62\% & 60\% &  &  & \\
				Autonomic & autonomous & autonomy  & & & \\ 
				& 70\% & 77\% & & & \\
				Self-configuration & self-configuring & configuration & configure & &  \\ 
				& 77\% & 72\% & 44\% & & \\
				Self-healing & self-recover & heal  & recover && \\
				& 50\% & 33\% & 16\% \\
				& repair & self-repair &&& \\ 
				& 25\% & 58\% &&&\\
				Self-optimization & self-improve & optimize & improve && \\
				& 46\% & 46\% & 13\% & & \\
				& performance  &&&& \\ 
				& 26\% &&&& \\
				Self-protection & self-protecting & protect  & & & \\ 
				&  86\% & 46\% & & & \\
				\hline
				\label{table:lexical}
			\end{tabular}
			\begin{tablenotes}\footnotesize
				\item[*] Lexical (Levenshtein) values are percentage of synonyms matching source string. 
			\end{tablenotes}
		\end{threeparttable}
	\end{table}

	From lexical similarity, we progressed to measuring semantic similarity or how similar the keywords are in meaning across the honey system literature sample (Table \ref{table:semantic}). Here, again we developed a Python program to measure cosine similarity between each keyword and a selected sentence in the sample paper. The output was a Mean of Means ($\mu x \bar{}$) across the sample for the keyword category and thus can be interpreted as the average percent of meaning similarity between a keyword and the sample of honey system literature. We also computed the Mean ($M$) for each keyword category in the control paper. 
	
	Overall, all autonomic computing keywords demonstrated a Mean of Means greater than $20\%$. In isolation, we might consider percentages less than $51\%$ to indicate low similarity. However, this is where having the control paper became important. For example, in the case of self-optimization, the results were identical between the sample and control ($49\%$). Self-protection had the largest difference between sample and control with a nine basis points difference. However, the remaining categories all demonstrated percentages within several deviations of one another.   
	
	\begin{table}[H]
		\centering
		\caption{Keyword Semantic Similarity}
		\begin{tabular}{l c c }
			\hline
			& Sample & Control \\ \hline \hline
			& $\mu x \bar{}$ & $M$ \\
			Self-adaptation & 0.30 & 0.28 \\
			Autonomic & 0.24 & 0.27 \\
			Self-configuration & 0.29 & 0.41 \\
			Self-healing & 0.27 & 0.30 \\
			Self-optimization & 0.49 & 0.49 \\
			Self-protection & 0.39 & 0.48\\
			\label{table:semantic}
		\end{tabular}
	\end{table}
	
	\section{Conclusion}
	
	Advanced persistent threats (APTs), ransomware, and zero-day exploits are becoming more sophisticated, which requires better security measures to safeguard enterprises. To develop innovative defense strategies, honeypots and honeynets can play a crucial role. These deceptive systems can help researchers and practitioners capture malicious behavior, enabling the creation of new detection mechanisms and a deeper understanding of evolving threats.
	
	Existing research has established that the deployment of honeypots and honeynets involves significant maintenance and overhead expenses. However, the challenges associated with honeypots and honeynets are not unique to these systems. The increasing complexity of modern computing has created a strong incentive to design and develop systems that can operate without human intervention. Such incentive has given rise the field of autonomic computing.
	
	State-of-the-art honey system research purports to include adjacent design and development principles. Terms such as adaptive, dynamic, intelligent, and learning are in common use. However, it is not clear to what extent honeypots and honeynets implement true autonomic computing principles. With that in mind, we posed a single research question: to what extent are autonomic computing principles expressed in forward edge honeypot and honeynet research?
	
	In summary, we found autonomic computing keywords present in the honey system literature sample. Therefore, there can be little doubt forward edge honeypot and honeynet literature exhibit markers of autonomic computing principles. Further, we conclude honeypots and honeynets are evolving from state-of-the-art \textit{self-adaptation} to forward edge \textit{autonomic computing} implementations. Furthermore, the synonym extraction and subsequent lexical similarity measurements corroborate autonomic keywords are used throughout the literature sample. Finally, the semantic analysis showed autonomic keywords have meaning consistent with autonomic computing principles. However, the case is not closed as these results do not paint the entire picture.
	
	We observed low frequencies of self-configuration, self-healing, and self-protection keywords. However, the few instances of such keywords were enough to generate semantic similarity on par with the control paper. Keep in mind, such comparison is from a group to an individual. Further, we found it interesting that self-optimization featured so prominently in both frequency analyses. We consider this notable because while the nine honey system papers in the self-adaptive category included some autonomic computing language, seven of the nine exhibited the term \textit{autonomic} and eight included \textit{self-optimization}. Likewise, the autonomic category papers demonstrated autonomic computing terms. Five of the six included \textit{autonomic} and four had the \textit{self-optimization} keyword. In both categories, the same honey system research showed few other keyword markers. Meanwhile, the autonomic computing control paper showed a more equalized distribution across the autonomic computing keywords.
	
	Overall, we feel there is sufficient data here to support future work in the convergence of autonomic computing and honeypot or honeynet deceptive systems. The results in this study could be extended by exploring the technical honeypot or honeynet implementation in the sample articles and testing each for autonomic behavior. Further, future work may be of interest in the design and implementation of individual autonomic computing principles in an exemplar honeypot. Both of these efforts would have significance for researchers and practitioners alike. Moreover, the comparison between honey system literature and seminal work in autonomic computing research might be interesting insofar as the work seeks to establish thresholds for autonomic computing principle implementation. This could be taken two ways. First, we might ask, how much of a honeypot's behavior must be autonomic for such as system to represent autonomic computing. Secondly, what ratio of the four autonomic computing principles must be present (i.e., three of the four) for a honeypot to exhibit autonomic behavior might be fruitful to investigate.

\bibliographystyle{unsrt}
\bibliography{references.bib}

\begin{thebibliography}{10}

\bibitem{Bringer_2012}
Matthew~L Bringer, Christopher~A Chelmecki, and Hiroshi Fujinoki.
\newblock A survey: Recent advances and future trends in honeypot research.
\newblock {\em International Journal of Computer Network and Information
  Security}, 4(10):63, 2012.

\bibitem{Campbell_2015}
Ronald~M Campbell, Keshnee Padayachee, and Themba Masombuka.
\newblock A survey of honeypot research: Trends and opportunities.
\newblock In {\em 2015 10th international conference for internet technology
  and secured transactions (ICITST)}, pages 208--212. IEEE, 2015.

\bibitem{Ikuomenisan_2022}
Gbenga Ikuomenisan and Yasser Morgan.
\newblock Meta-review of recent and landmark honeypot research and surveys.
\newblock {\em Journal of Information Security}, 13(4):181--209, 2022.

\bibitem{Gupta_2018}
Brij~B Gupta and Alisha Gupta.
\newblock Assessment of honeypots: Issues, challenges and future directions.
\newblock {\em International Journal of Cloud Applications and Computing
  (IJCAC)}, 8(1):21--54, 2018.

\bibitem{horn2001autonomic}
Paul Horn.
\newblock Autonomic computing: Ibm’s perspective on the state of information
  technology.
\newblock 2001.

\bibitem{psaier2011survey}
Harald Psaier and Schahram Dustdar.
\newblock A survey on self-healing systems: approaches and systems.
\newblock {\em Computing}, 91:43--73, 2011.

\bibitem{choudhury2021autonomic}
Tanupriya Choudhury, Bhupesh~Kumar Dewangan, Ravi Tomar, Bhupesh~Kumar Singh,
  Teoh~Teik Toe, and Nguyen~Gia Nhu.
\newblock {\em Autonomic Computing in Cloud Resource Management in Industry
  4.0}.
\newblock Springer, 2021.

\bibitem{Parashar2004}
Manish Parashar and Salim Hariri.
\newblock Parashar, manish, and salim hariri.
\newblock In {\em In International workshop on unconventional programming
  paradigms}, pages 257--269. Springer, 2004.

\bibitem{Dehraj2021}
Pooja Dehraj and Arun Sharma.
\newblock A review on architecture and models for autonomic software systems.
\newblock {\em The Journal of Supercomputing}, 77(1):388--417, 2021.

\bibitem{kephart2003}
Jeffrey~O Kephart and David~M Chess.
\newblock The vision of autonomic computing.
\newblock {\em Computer}, 36(1):41--50, 2003.

\bibitem{Henrichs2022}
Elia Henrichs, Veronika Leschb, Martin Straesserb, Samuel Kounevb, and
  Christian Krupitzera.
\newblock A literature review on optimization techniques for adaptation
  planning in adaptive systems: State of the art and research directions.
\newblock {\em Information and Software Technology}, 2022.

\bibitem{Quin2022}
Federico Quin, Danny Weyns, and Omid Gheibi.
\newblock Reducing large adaptation spaces in self-adaptive systems using
  classical machine learning.
\newblock {\em The Journal of Systems and Software}, 190:111341, 2022.

\bibitem{Wong2022}
Terence Wong; Markus Wagner;~Christoph Treude;.
\newblock Self-adaptive systems: A systematic literature review across
  categories and domains.
\newblock {\em Information and Software Technology}, 148(2022), 2022.

\bibitem{curran2005monitoring}
Kevin Curran, Colman Morrissey, Colm Fagan, Colm Murphy, Brian O'Donnell, Gerry
  Fitzpatrick, and Stephen Condit.
\newblock Monitoring hacker activity with a honeynet.
\newblock {\em International Journal of Network Management}, 15(2):123--134,
  2005.

\bibitem{Zakaria_2013}
Miss Laiha Mat~Kiah Wira Zanoramy Ansiry~Zakaria.
\newblock A review of dynamic and intelligent honeypots.
\newblock 2013.

\bibitem{Mohan_2022}
Pilla~Vaishno Mohan, Shriniket Dixit, Amogh Gyaneshwar, Utkarsh Chadha,
  Kathiravan Srinivasan, and Jung~Taek Seo.
\newblock Leveraging computational intelligence techniques for defensive
  deception: a review, recent advances, open problems and future directions.
\newblock {\em Sensors}, 22(6):2194, 2022.
\newblock cites landsborough2021towards.

\bibitem{Mohammadzadeh_2013}
Hamid Mohammadzadeh, Masood Mansoori, and Ian Welch.
\newblock Evaluation of fingerprinting techniques and a windows-based dynamic
  honeypot.
\newblock In {\em Proceedings of the Eleventh Australasian Information Security
  Conference-Volume 138}, pages 59--66, 2013.

\bibitem{Han_2016}
Wonkyu Han, Ziming Zhao, Adam Doup{\'e}, and Gail-Joon Ahn.
\newblock Honeymix: Toward sdn-based intelligent honeynet.
\newblock In {\em Proceedings of the 2016 ACM International Workshop on
  Security in Software Defined Networks \& Network Function Virtualization},
  pages 1--6, 2016.

\bibitem{Hecker2012}
Christopher~R Hecker.
\newblock {\em A methodology for intelligent honeypot deployment and active
  engagement of attackers}.
\newblock PhD thesis, 2012.

\bibitem{Fan2019}
Wenjun Fan, Zhihui Du, Max Smith-Creasey, and David Fernandez.
\newblock Honeydoc: an efficient honeypot architecture enabling all-round
  design.
\newblock {\em IEEE Journal on Selected Areas in Communications},
  37(3):683--697, 2019.

\bibitem{Rowe2019}
Neil~C Rowe.
\newblock Honeypot deception tactics.
\newblock {\em Autonomous cyber deception: Reasoning, adaptive planning, and
  evaluation of HoneyThings}, pages 35--45, 2019.

\bibitem{Pittman2020}
Jason~M Pittman, Kyle Hoffpauir, Nathan Markle, and Cameron Meadows.
\newblock A taxonomy for dynamic honeypot measures of effectiveness.
\newblock {\em arXiv preprint arXiv:2005.12969}, 2020.

\bibitem{Kuwatly_2004}
Iyad Kuwatly, Malek Sraj, Zaid Al~Masri, and Hassan Artail.
\newblock A dynamic honeypot design for intrusion detection.
\newblock In {\em The IEEE/ACS International Conference onPervasive Services,
  2004. ICPS 2004. Proceedings.}, pages 95--104. IEEE, 2004.

\bibitem{Hecker_2006}
Christopher Hecker, Kara~L Nance, and Brian Hay.
\newblock Dynamic honeypot construction.
\newblock In {\em Proceedings of the 10th Colloquium for Information Systems
  Security Education}, volume 102. Citeseer, 2006.

\bibitem{Park_2019}
Byungju Park, Sa~Pham Dang, Sichul Noh, Junmin Yi, and Minho Park.
\newblock Dynamic virtual network honeypot.
\newblock In {\em 2019 International Conference on Information and
  Communication Technology Convergence (ICTC)}, pages 375--377. IEEE, 2019.

\bibitem{Mitchell2018}
Aidan Mitchell.
\newblock An intelligent honeypot.
\newblock {\em Cork Institute of Technology}, 2018.

\bibitem{Meng_2017}
Xiangjun Meng, Zhifeng Zhao, Rongpeng Li, and Honggang Zhang.
\newblock An intelligent honeynet architecture based on software defined
  security.
\newblock In {\em 2017 9th International Conference on Wireless Communications
  and Signal Processing (WCSP)}, pages 1--6. IEEE, 2017.

\bibitem{Naik_2020}
Nitin Naik, Paul Jenkins, Nick Savage, and Longzhi Yang.
\newblock A computational intelligence enabled honeypot for chasing ghosts in
  the wires.
\newblock {\em Complex \& Intelligent Systems}, 7:477--494, 2021.

\bibitem{Chakraborty2016}
Manajit Chakraborty, Sukomal Pal, Rahul Pramanik, and C~Ravindranath Chowdary.
\newblock Recent developments in social spam detection and combating
  techniques: A survey.
\newblock {\em Information Processing \& Management}, 52(6):1053--1073, 2016.

\bibitem{Jiang_2020}
Kui Jiang and Haocheng Zheng.
\newblock Design and implementation of a machine learning enhanced web honeypot
  system.
\newblock In {\em 2020 13th International Congress on Image and Signal
  Processing, BioMedical Engineering and Informatics (CISP-BMEI)}, pages
  957--961. IEEE, 2020.

\bibitem{MELLIA2021}
MARCO MELLIA.
\newblock {\em Cannypot: A Reinforcement Learning Based Adaptive SSH Honeypot}.
\newblock PhD thesis, Politecnico di Torino, 2021.

\bibitem{Fan_2015}
Wenjun Fan, David Fern{\'a}ndez, and Zhihui Du.
\newblock Adaptive and flexible virtual honeynet.
\newblock In {\em International Conference on Mobile, Secure and Programmable
  Networking}, pages 1--17. Springer, 2015.

\bibitem{Dowling_2020}
Seamus Dowling, Michael Schukat, and Enda Barrett.
\newblock New framework for adaptive and agile honeypots.
\newblock {\em Etri Journal}, 42(6):965--975, 2020.

\bibitem{Landsborough2021}
Jason Landsborough, Luke Carpenter, Braulio Coronado, Sunny Fugate, Kimberly
  Ferguson-Walter, and Dirk Van~Bruggen.
\newblock Towards self-adaptive cyber deception for defense.
\newblock In {\em HICSS}, pages 1--10, 2021.

\bibitem{Zhang_2017}
Yan Zhang, Chong Di, Zhuoran Han, Yichen Li, and Shenghong Li.
\newblock An adaptive honeypot deployment algorithm based on learning automata.
\newblock In {\em 2017 IEEE Second International Conference on Data Science in
  Cyberspace (DSC)}, pages 521--527. IEEE, 2017.

\bibitem{Wagener_2009}
G{\'e}rard Wagener, DA~Radu~State, Alexandre Dulaunoy, and Thomas Engel.
\newblock Self adaptive high interaction honeypots driven by game theory.
\newblock In {\em SSS}, pages 741--755. Springer, 2009.

\bibitem{Touch_2022}
Sereysethy Touch and Jean-No{\"e}l Colin.
\newblock A comparison of an adaptive self-guarded honeypot with conventional
  honeypots.
\newblock {\em Applied Sciences}, 12(10):5224, 2022.

\bibitem{Fraunholz_2017}
Daniel Fraunholz, Marc Zimmermann, Alexander Hafner, and Hans~D Schotten.
\newblock Data mining in long-term honeypot data.
\newblock In {\em 2017 IEEE International Conference on Data Mining Workshops
  (ICDMW)}, pages 649--656. IEEE, 2017.

\bibitem{Huang_2019}
Linan Huang and Quanyan Zhu.
\newblock Adaptive honeypot engagement through reinforcement learning of
  semi-markov decision processes.
\newblock 1906.

\bibitem{Petticrew2008}
Mark Petticrew and Helen Roberts.
\newblock {\em Systematic reviews in the social sciences: A practical guide}.
\newblock John Wiley \& Sons, 2008.

\bibitem{Cooper1998}
Harris~M Cooper.
\newblock {\em Synthesizing research: A guide for literature reviews},
  volume~2.
\newblock Sage, 1998.

\bibitem{Kitchenham2004}
Barbara Kitchenham.
\newblock Procedures for performing systematic reviews.
\newblock {\em Keele, UK, Keele University}, 33(2004):1--26, 2004.

\bibitem{DeFaveri2016}
Cristiano De~Faveri and Ana Moreira.
\newblock Designing adaptive deception strategies.
\newblock In {\em 2016 IEEE International Conference on Software Quality,
  Reliability and Security Companion (QRS-C)}, pages 77--84. IEEE, 2016.

\bibitem{Kashtalian2021}
Antonina Kashtalian and Tomas Sochor.
\newblock K-means clustering of honeynet data with unsupervised representation
  learning.
\newblock In {\em IntelITSIS}, pages 439--449, 2021.

\bibitem{Niakanlahiji2020}
Amirreza Niakanlahiji, Jafar~Haadi Jafarian, Bei-Tseng Chu, and Ehab Al-Shaer.
\newblock Honeybug: Personalized cyber deception for web applications.
\newblock 2020.

\bibitem{Pauna2018}
Adrian Pauna, Andrei-Constantin Iacob, and Ion Bica.
\newblock Qrassh-a self-adaptive ssh honeypot driven by q-learning.
\newblock In {\em 2018 international conference on communications (COMM)},
  pages 441--446. IEEE, 2018.

\bibitem{Sun2022}
Chongxin Sun, Youjun Bu, Bo~Chen, Desheng Zhang, Zhonglei Chen, Xiangyu Lu,
  Surong Zhang, and Jia Sun.
\newblock Application of artificial intelligence technology in honeypot
  technology.
\newblock In {\em 2021 International Conference on Advanced Computing and
  Endogenous Security}, pages 01--09. IEEE, 2022.

\bibitem{Suratkar2022}
Shraddha Suratkar, Kunjal Shah, Aditya Sood, Anay Loya, Dhaval Bisure, Umesh
  Patil, and Faruk Kazi.
\newblock An adaptive honeypot using q-learning with severity analyzer.
\newblock {\em Journal of Ambient Intelligence and Humanized Computing},
  13(10):4865--4876, 2022.

\bibitem{feeneyhoneynet}
Mairead Feeney and Kevin Curran.
\newblock Honeynet intelligent network behaviour analysis.

\bibitem{Li2019}
Yan Li and Bin Wu.
\newblock Design and implementation of modular honeynet system based on sdn.
\newblock In {\em International Conference on Advances in Computer Technology,
  Information Science and Communications}, 2019.

\bibitem{li2009experiment}
Bingyang Li, Huiqiang Wang, and Guangsheng Feng.
\newblock Experiment research of automatic deception model based on autonomic
  computing.
\newblock In {\em Advances in Information Security and Its Application: Third
  International Conference, ISA 2009, Seoul, Korea, June 25-27, 2009.
  Proceedings 3}, pages 98--104. Springer, 2009.

\bibitem{teles2011autonomic}
Ariel~S Teles, Jean~PM Mendes, and Zair Abdelouahab.
\newblock Autonomic computing applied to network security: A survey.
\newblock {\em Cyber Journals: Multidisciplinary Journals in Science and
  Technology, Journal of Selected Areas in Telecommunications (JSAT), November
  Edition}, pages 7--14, 2011.

\bibitem{Zarca2020}
Alejandro~Molina Zarca, Jorge~Bernal Bernabe, Antonio Skarmeta, and Jose
  M~Alcaraz Calero.
\newblock Virtual iot honeynets to mitigate cyberattacks in sdn/nfv-enabled iot
  networks.
\newblock {\em IEEE Journal on Selected Areas in Communications},
  38(6):1262--1277, 2020.

\bibitem{You2021}
Jianzhou You, Shichao Lv, Yue Sun, Hui Wen, and Limin Sun.
\newblock Honeyvp: A cost-effective hybrid honeypot architecture for industrial
  control systems.
\newblock In {\em ICC 2021-IEEE International Conference on Communications},
  pages 1--6. IEEE, 2021.

\bibitem{Pennington2014}
Jeffrey Pennington, Richard Socher, and Christopher~D Manning.
\newblock Glove: Global vectors for word representation.
\newblock In {\em Proceedings of the 2014 conference on empirical methods in
  natural language processing (EMNLP)}, pages 1532--1543, 2014.

\end{thebibliography}

\end{document}